# Local unitary symmetries of hypergraph states

David W. Lyons[1], Daniel J. Upchurch[2], Scott N. Walck[3], Chase D. Yetter[4]
Departments of Mathematical Sciences and Physics
Lebanon Valley College

**Abstract**

Hypergraph states are multiqubit states whose combinatorial description and entanglement properties generalize the well-studied class of graph states. Graph states are important in applications such as measurement-based quantum computation and quantum error correction. The study of hypergraph states, with their richer multipartite entanglement and other nonlocal properties, has a promising outlook for new insight into multipartite entanglement. We present results analyzing local unitary symmetries of hypergraph states, including both continuous and discrete families of symmetries. In particular, we show how entanglement types can be detected and distinguished by certain configurations in the hypergraphs from which hypergraph states are constructed.

## 1 Introduction

The idea of harnessing quantum systems to perform computational tasks, suggested by Feynman in 1982 [1] and propelled to mainstream attention in physics, mathematics, and computer science with Peter Shor's discovery of a quantum factoring algorithm in 1994 [2], still faces engineering hurdles to full-scale practical implementation. In the meantime, theoreticians have developed schemes for quantum computation and error correction that will, in principle, carry out quantum algorithms while protecting against noise and decoherence. Important among these methods is the so-called *1-way quantum computation* that exploits multiqubit states called *graph states* as resources in computational protocols [3].

An appealing feature of graph states is their conceptually simple construction: given a graph on $n$ nodes or vertices, one prepares a product state of $n$ qubits, and then one applies a 2-qubit entangling operation to pairs of the qubits as specified by the edges of the given graph. All of these 2-qubit gates are the same (the controlled-$Z$ gate, described below). Remarkably, this simple recipe is enough to create resource states rich enough for universal quantum computation and sufficiently robust encoding to implement quantum algorithms subject to realistic noise models [3].

A number of recent papers [4, 5, 6, 7, 8, 9] consider a natural generalization of graph states called *hypergraph* states. As with graph states, there is an underlying combinatorial object—the hypergraph—and construction rules for turning hypergraphs into multiqubit quantum states requiring only one kind of multiparty entangling gate (the *generalized* controlled-$Z$ gates). The simplicity of graph and hypergraph states supports plausibility arguments for experimental realization; an entanglement purification protocol that produces hypergraph states (Carle et al. [10]) provides concrete evidence supporting this view. At the same time, graph and hypergraph states have rich nonlocal properties. The coding theory of graph states has led to the development of the so-called *stabilizer formalism*—the theory of the discrete Pauli tensor group [11]. Recent work of Makmal et al. [12], who show that the NP-complete SAT problem can be solved efficiently if one has access to a black box that can determine whether an input hypergraph state is a product, demonstrates the nontriviality of hypergraph states.

[1]lyons@lvc.edu, Department of Mathematical Sciences
[2]dju001@lvc.edu, Department of Physics
[3]walck@lvc.edu, Department of Physics
[4]cdy001@lvc.edu, Department of Mathematical Sciences

In this paper, we study entanglement properties of hypergraph states. A basic question at the heart of understanding entanglement is that of local unitary (LU) equivalence: given hypergraph states arising from different hypergraphs, can one be transformed to the other by local unitary operations? A special case, already nontrivial, is whether a given hypergraph state can be transformed by local unitary operations into a graph state. More generally: what, precisely, are the equivalence classes of hypergraph states under the action of the local unitary group?

While there are solutions to the problem of LU equivalence, many questions of entanglement classification and operational meaning remain. Kraus[13, 14] and Martins [15] give criteria and describe algorithms which determine whether two input states are LU equivalent. However, these methods do not provide a classification of LU equivalence classes, nor do they address operational questions of entanglement properties of states. Recent work of Gühne et al. [8] establishes the beautiful result that for a generic class of hypergraph states (that also meet some specific criteria), two states are local unitary equivalent if and only if they are local Pauli equivalent. This is striking in light of the history of the "LU-LC conjecture" which says that two *graph* states are local unitary equivalent if and only if they are local Clifford equivalent. The subtlety of the failure of the LU-LC conjecture is reflected in the nontriviality of the counterexamples [16]. Thus there is reason to expect that the LU equivalence story for hypergraphs will not be a simple one.

This paper presents several avenues of investigation of the local unitary action on hypergraph states. While we do not claim an exhaustive analysis, we develop tools and establish results that add to what is known and point to future investigations. After preliminary sections 2 and 3, we give results on the LU action on and LU symmetries of hypergraph states in sections 3 and 4. We examine discrete symmetries of permutationally invariant hypergraph states in section 6, and describe density matrices of subsystems of hypergraph states in section 7, addressing the problem of reconstructing an $n$-qubit hypergraph state from a given reduced density matrix.

## 2 Hypergraphs and hypergraph states

A hypergraph is a pair $G = (V, E)$ where $V$ is a finite set of *vertices* and $E$ is a collection of subsets of $V$ called the *hyperedges* of $G$. Given a hypergraph $G$ with $n$ vertices labeled $1, 2, \ldots, n$, we define the corresponding $n$-qubit quantum hypergraph state $|\psi_G\rangle$ by

$$|\psi_G\rangle := \left(\prod_{e\in E} C_e\right) |+\rangle^{\otimes n} \tag{1}$$

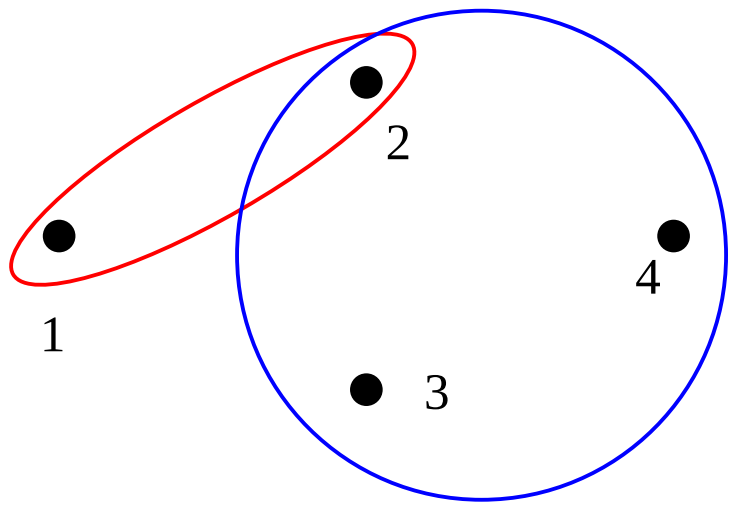

$$\begin{aligned}|\psi\rangle &= C_{12}C_{234}\,|+\rangle^{\otimes 4}\\ &= |0000\rangle+|0001\rangle+|0010\rangle+|0011\rangle+|0100\rangle\\ &+ |0101\rangle+|0110\rangle+|1000\rangle+|1001\rangle+|1010\rangle+|1011\rangle\\ &- |0111\rangle-|1100\rangle-|1101\rangle-|1110\rangle+|1111\rangle\end{aligned}$$

Figure 1: Example: a hypergraph and its corresponding state. Red and blue (color version) match the hyperedges, the $C_e$ gates, and the ket vectors on which they act. Both gates act on qubit 2 in the last term $|1111\rangle$.

where $C_e$ denotes the *generalized controlled-$Z$ operator*, defined by its action on the computational basis vector $|I\rangle = |i_1 i_2 \cdots i_n\rangle$ (each $i_k = 0, 1$) as follows.

$$C_e\,|I\rangle := (-1)^{\prod_{k\in e} i_k}\,|I\rangle \tag{2}$$

We will use the notation $\wedge_e I$ to denote the product $\prod_{k\in e} i_k$ (thinking of multiplying 0's and 1's as the logical 'and' operation on the bits of $I$ in the subset given by $e$), so that (2) has the more compact form

$$C_e\,|I\rangle = (-1)^{\wedge_e I}\,|I\rangle\,. \tag{3}$$

Said another way, $C_e$ applies the operator

$$\mathbb{1} - 2(|1\rangle\langle 1|)^{\otimes|e|} \tag{4}$$

to qubits in the hyperedge $e$, and applies the identity on remaining qubits. See Figure 1 for an example.

For a hyperedge with only a few vertices, say $e = \{a, b, c\}$, we will write $C_{abc}$ instead of the more cumbersome $C_{\{a,b,c\}}$. In particular, $C_k$ is the same as the 1-qubit operator $Z_k$, and $C_{ab}$ is the ordinary controlled-$Z$ operating on qubits $a, b$ (hence the name "generalized controlled-$Z$").

We say a hypergraph or its corresponding state is $m$-uniform if all hyperedges contain exactly $m$ vertices. In this language, a graph is a 2-uniform hypergraph for which all of the $C_e$ operators are ordinary controlled-$Z$ gates.

It is convenient to introduce notation for computational basis vectors that have 0's and 1's in specified positions. Given a subset $S \subseteq \{1, 2, \ldots, n\}$ of vertices, we write $|1_S\rangle$ to denote the computational basis vector that has 1's in positions given by $S$ and 0's elsewhere (for the special case when $|S| = 1$, say $S = \{b\}$, we write $|1_b\rangle$ instead of the more cumbersome $|1_{\{b\}}\rangle$). The utility of this notation derives from the following elementary observation.

$$C_e\,|1_S\rangle = \begin{cases} -|1_S\rangle & \text{if } e \subseteq S \\ \;\;|1_S\rangle & \text{otherwise} \end{cases} \tag{5}$$

We conclude this section with two useful expressions for $|\psi_G\rangle$. Using

$$\sqrt{2^n}\,|+\rangle^{\otimes n} = \sum_I |I\rangle = \sum_{S\subseteq\{1,\ldots,n\}} |1_S\rangle \tag{6}$$

we apply (3) and (5), respectively, to the definition (1) to obtain the following.

$$\begin{aligned} \sqrt{2^n}\,|\psi_G\rangle &= \sum_I (-1)^{\sum_{e\in E}\wedge_e I}\,|I\rangle & (7)\\ &= \sum_{S\subseteq\{1,\ldots n\}} (-1)^{\#\{e\in E\colon\, e\subseteq S\}}\,|1_S\rangle & (8)\end{aligned}$$

# 3 The local unitary group and algebra

Let $G = U(1)\times \underbrace{SU(2)\times\cdots\times SU(2)}_{n \text{ factors}}$ denote the local unitary (LU) group[5] for $n$-qubit states, where an element $g = (e^{it}, g_1, \ldots, g_n)$ in $G$ acts on an $n$-qubit pure state $|\psi\rangle$ by

$$|\psi\rangle \to e^{it} g_1\otimes\cdots\otimes g_n\,|\psi\rangle\,.$$

Let $\mathfrak{g} = u(1)\oplus su(2)\oplus\ldots\oplus su(2)$ denote the Lie algebra of $G$. Here, $u(1)$ is the set of imaginary numbers (the Lie algebra of the group $U(1)$), and $su(2)$ is the set of $2\times 2$ skew-hermitian matrices (the Lie algebra of the group $SU(2)$). For $su(2)$ we use the basis $iX, iY, iZ$ ($i$ times the Pauli matrices $X, Y, Z$), and we write

$$i\left(\theta + \sum_{k=1}^{n}(r_k X_k + s_k Y_k + t_k Z_k)\right) \tag{9}$$

to denote a general element of $\mathfrak{g}$, where $\theta, r_k, s_k, t_k$ are real and $X_k, Y_k, Z_k$ denote Pauli $X, Y, Z$ (respectively) acting in the $k$th position.

Given a state $|\psi\rangle$, we write $\mathrm{Stab}_\psi$ to denote the LU stabilizer subgroup (also called the *LU symmetry group*)

$$\mathrm{Stab}_\psi = \{g\in G\colon\, g\,|\psi\rangle = |\psi\rangle\}$$

and we write $K_\psi$ to denote the LU stabilizer subalgebra

$$K_\psi = \{M\in\mathfrak{g}\colon\, M\,|\psi\rangle = 0\}$$

which is the same as the Lie algebra of the group $\mathrm{Stab}_\psi$. [Note: Solving equations of the form $M\,|\psi\rangle = 0$, it does not matter whether we consider skew-hermitian $M\in K_\psi$ or hermitian $\frac{1}{i}M$. For this reason, we omit a factor of $i$ in LU stabilizer algebra calculations when convenient.] The *weight* of an LU group element $g = (e^{it}, g_1, \ldots, g_n)$ or an LU algebra element $M = (t, M_1, \ldots, M_n)$ is the number of the $g_k$ not proportional to the identity or the number of the $M_k$ that are nonzero, respectively.

Our chief motivation for studying LU symmetry groups is their role in LU equivalence and entanglement. If $|\psi\rangle, |\psi'\rangle$ are LU equivalent, say $|\psi'\rangle = g\,|\psi\rangle$

[5] Context will ensure that the symbol '$G$', used in this section to denote the local unitary group, will not be confused for a hypergraph in other sections.

for some LU transformation $g$, then their LU stabilizer groups and algebras are isomorphic via conjugation [17].

$$\mathrm{Stab}_{\psi'} = g\mathrm{Stab}_{\psi}g^{\dagger} \tag{10}$$

$$K_{\psi'} = gK_{\psi}g^{\dagger} \tag{11}$$

Thus (conjugacy classes of) symmetry groups and algebras constitute LU invariants. If symmetry groups of two states differ in some property that is preserved by conjugation, for example dimension, then the two states are *not* LU equivalent. In this way, symmetry groups provide a means of course graining the space of entanglement classes.

# 4 LU algebra action on hypergraph states

In this section we extend results of Zhang, Fan, and Zhou in their paper [18] to hypergraph states. The proof is in the appendix.

**(4.1) Proposition.** *Let $|\psi_G\rangle$ denote an $n$-qubit hypergraph state with hypergraph $G = (V, E)$ with vertex set $V$ and hyperedge set $E$. The actions of $X_a, Y_a$ on $|\psi_G\rangle$ are given by the following.*

$$X_a |\psi_G\rangle = \prod_{e\colon a\in e} C_{e\setminus a} |\psi_G\rangle \tag{12}$$

$$Y_a |\psi_G\rangle = -iZ_a \prod_{e\colon a\in e} C_{e\setminus a} |\psi_G\rangle \tag{13}$$

It follows that the element

$$M = \theta + \sum_{a\in V} (r_a X_a + s_a Y_a + t_a Z_a) \tag{14}$$

acts on $|\psi_G\rangle$ by

$$|\psi_G\rangle \to \Phi_M |\psi_G\rangle . \tag{15}$$

where the operator $\Phi_M$ is given by

$$\Phi_M := \theta + \sum_{a\in V} \left( r_a \prod_{e\colon a\in e} C_{e\setminus a} - is_a Z_a \prod_{e\colon a\in e} C_{e\setminus a} + t_a Z_a \right) . \tag{16}$$

For the sake of compactness, in what follows we write $P_a$ to denote

$$P_a = \prod_{e\colon a\in e} C_{e\setminus a} \tag{17}$$

so that equation (16) reads

$$\Phi_M := \theta + \sum_{a\in V} (r_a P_a - is_a Z_a P_a + t_a Z_a) . \tag{18}$$

## 5 LU stabilizer algebra for hypergraph states

In this section we apply the Lie algebra action formulas of the previous section to establish results about the stabilizer subalgebra $K_{\psi_G}$ of a hypergraph state $|\psi_G\rangle$. Later in the section we compare stabilizers for general hypergraph states to the special case of graph states (graph states are hypergraphs for which all hyperedges contain exactly two vertices) and draw some conclusions about LU equivalence classes.

We begin with the observation that because $|\psi_G\rangle$ is a superposition of all the computational basis vectors (with coefficients $\pm 1/\sqrt{2^n}$) and $\Phi_M$ acts diagonally in this basis, we have $\Phi_M |\psi_G\rangle = 0$ if and only if $\Phi_M = 0$. Separating real and imaginary parts of (18), we have the following.

**(5.2) Proposition.** *Let $G = (V, E)$ be a hypergraph and let $|\psi_G\rangle$ be the corresponding hypergraph state. The element $M = \theta + \sum_{a \in V} (r_a X_a + s_a Y_a + t_a Z_a)$ satisfies $M |\psi_G\rangle = 0$, that is, $iM \in K_{\psi_G}$, if and only if the following two equations hold.*

$$\theta + \sum_{a \in V} (r_a P_a + t_a Z_a) = 0 \tag{19}$$

$$\sum_{a \in V} s_a Z_a P_a = 0 \tag{20}$$

We now apply (5.2) to hypergraph states for which all hyperedges have at least three vertices. (The proof is in the appendix.)

**(5.3) Theorem.** *Let $G$ be a hypergraph such that all hyperedges contain at least three vertices and let $|\psi_G\rangle$ be the corresponding hypergraph state. Then every element in the stabilizer Lie subalgebra $K_{\psi_G}$ is of the form*

$$i\left(\theta + \sum_k r_k X_k\right)$$

*for some real $\theta, r_k$.*

Given a qubit $a$ and a hyperedge $e$ containing $a$, we call $e \setminus \{a\}$ a ***reduced set*** for $a$. We say a subset $S$ of the vertex set is a ***shared reduced set*** if it is the reduced set for more than one vertex. The next proposition gives a criterion in terms of reduced sets for when $r_a$ must be zero (the proof is in the appendix).

**(5.4) Proposition.** *Let $G$ be an $m$-uniform hypergraph for some $m \geq 3$, with corresponding hypergraph state $|\psi_G\rangle$. If a qubit $b$ has a reduced set that is not shared then $r_b = 0$.*

The previous proposition says that in order for an $m$-uniform hypergraph state to have a nontrivial stabilizer algebra, there must be at least two qubits that share reduced sets, that is, there must be some qubits $a, b$ with hyperedges $e$ containing $a$ and $f$ containing $b$ with $e \setminus \{a\} = f \setminus \{b\}$ (see Figure 2). This motivates the following definitions.

Given an $m$-uniform hypergraph $G = (V, E)$ with $m \geq 3$, we define the ***essential vertices*** to be the set $\hat{V} \subseteq V$ of vertices $v$ in $V$ for which every reduced set is shared with some other vertex. We define the ***essential hyperedges*** to be the subsets of essential vertices, maximal with respect to the property that every vertex in the hyperedge shares a reduced set with every other vertex in the hyperedge. Let $\hat{E}$ denote the set of essential hyperedges. Finally, we define the ***essential hypergraph*** for $G$ to be the hypergraph $\hat{G} = (\hat{V}, \hat{E})$. See Figure 3. The term "essential" is warranted in the sense that if a hypergraph state (such that each hyperedge contains 3 or more vertices) has an empty essential hypergraph, then that state has no nontrivial LU stabilizer algebra elements, and thus has at most only discrete LU group symmetries.

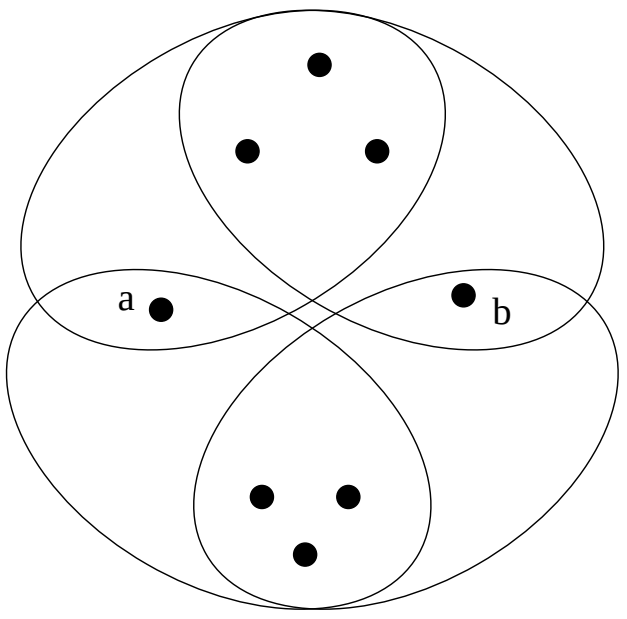

Figure 2: Qubits $a, b$ share reduced sets, and $P_a = P_b$.

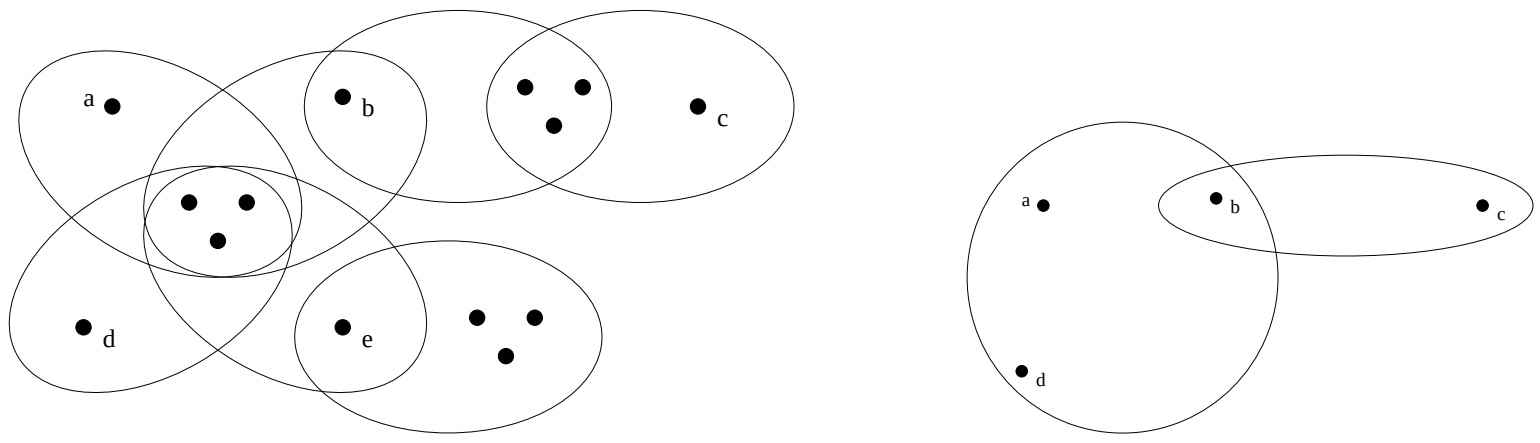

Figure 3: A hypergraph (left) and its essential hypergraph (right). Unlabeled vertices are not essential because they have no shared reduced sets. Vertex $e$ is not essential because it has a reduced set that is not shared.

Here is an example of a stabilizer result that has a natural phrasing in terms of the essential hypergraph. (The proof is in the appendix.)

**(5.5) Proposition.** *Suppose that $|\psi_G\rangle$ is an $m$-uniform hypergraph state for some $m \geq 3$, and for which the essential hypergraph $\hat{G} = (\hat{V}, \hat{E})$ has exactly one hyperedge. Then $K_{\psi_G}$ has dimension $|\hat{V}|-1$, with basis $i(X_1 - X_j)$, $2 \leq j \leq |\hat{V}|$, where $1, 2, \ldots, |\hat{V}|$ are the qubit labels for the essential vertices.*

The assumption that there is one essential hyperedge implies that the underlying hypergraph $G$ has the configuration shown in Figure 4 (see the proof in the appendix). This family of states demonstrates the existence of high dimensional LU symmetry groups for hypergraph states.

Here is another result about essential hypergraphs. It demonstrates that, while having a nonempty essential hypergraph is a *necessary* condition for the existence of a nonzero LU stabilizer algebra (for hypergraph states with hyperedges all size 3 or more), it is not a *sufficient* condition. (The proof is in the appendix.)

**(5.6) Proposition.** *Let $|\psi_G\rangle$ be a hypergraph state with hyperedges size 3 or more. If the essential hypergraph $\hat{G}$ has at least 3 essential vertices, is 2-uniform and connected (that is, $\hat{G}$ is a connected graph), and all reduced sets are shared*

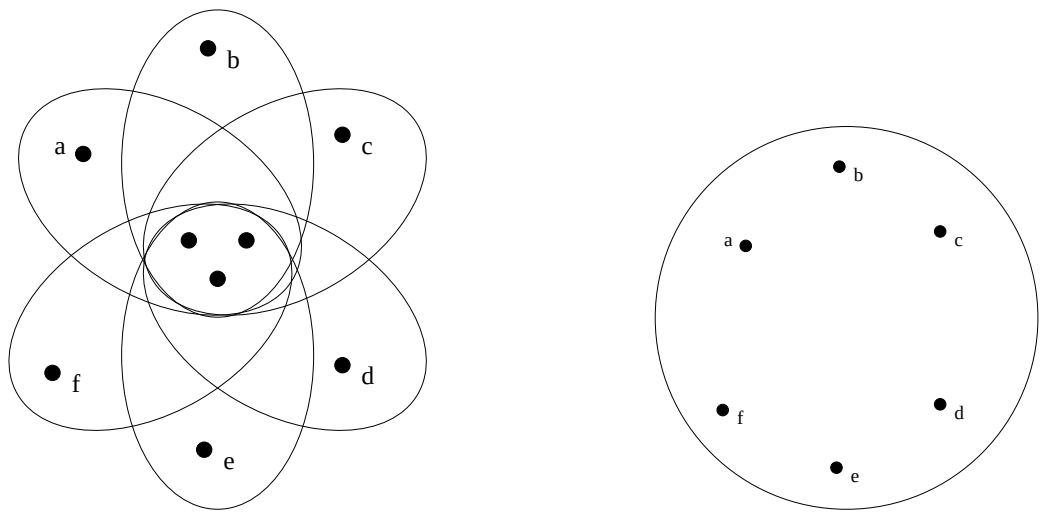

Figure 4: A hypergraph (left) and its essential hypergraph (right) with high dimensional LU symmetry group.

*by at most two essential vertices, then* $K_{\psi_G} = 0$.

We continue with a comparison to the case of graph states, that is, the subclass of hypergraph states for which all hyperedges contain exactly two vertices. Zhang et al. [18] show that the local unitary stabilizer algebra for any graph state is generated by weight 2 elements of the form $X_a - X_b$, $Y_a - Y_b$, and $X_a - Z_b$, which correspond to certain configurations in the underlying graph. By contrast, we will show that hypergraph states can have stabilizer algebras with generators of arbitrarily high weight. These weight considerations are interesting because minimal generator weights can distinguish LU equivalence classes and therefore entanglement classes (because the weight of an LU group element or an LU algebra element is preserved by LU conjugation, the lowest set of weights among all sets of generators of the LU symmetry group (or basis elements of the LU stabilizer subalgebra) is an LU invariant).

Here is an example. Consider generalized controlled-$Z$ gates in two qubits: $C_1, C_2, C_{1,2}$. It is elementary to check that the following linear relation among their products holds.

$$-C_1 - C_2 + C_{1,2} + C_1 C_2 C_{1,2} = 0. \tag{21}$$

By adding extra qubits and choosing appropriate hyperedges, we can realize (21) as the operator $\Phi_M$ for an appropriate $M$. Consider the vertex set $V = \{1, 2, a, b, c, d\}$, let $E$ be the collection of hyperedges

$$\{a, 1\}, \{b, 2\}, \{c, 1, 2\}, \{d, 1\}, \{d, 2\}, \{d, 1, 2\}$$

and let $G = (V, E)$ (see Figure 5). These choices construct a hypergraph state $|\psi_G\rangle$ such that the four operators $P_a, P_b, P_c, P_d$ are equal to the four terms $C_1, C_2, C_{1,2}, C_1C_2C_{1,2}$ in (21). Thus, for

$$M = -X_a - X_b + X_c + X_d \tag{22}$$

we have $\Phi_M$ equal to the left side of (21), so $iM$ is an element of $K_{\psi_G}$. It is an elementary check (if tedious by hand) to show that $K_{\psi_G}$ is 1-dimensional, so it has no generators of weight less than 4. We immediately conclude that this hypergraph state cannot be LU equivalent to any graph state.

This construction generalizes as follows. Let $\mathcal{P}(\{1, \ldots, m\})$ denote the power set of $\{1, 2, \ldots, m\}$ and suppose there is a nontrivial linear relation

$$\sum_{S \subseteq \mathcal{P}(\{1,\ldots,m\})} c_S \prod_{e \in S} C_e = 0 \tag{23}$$

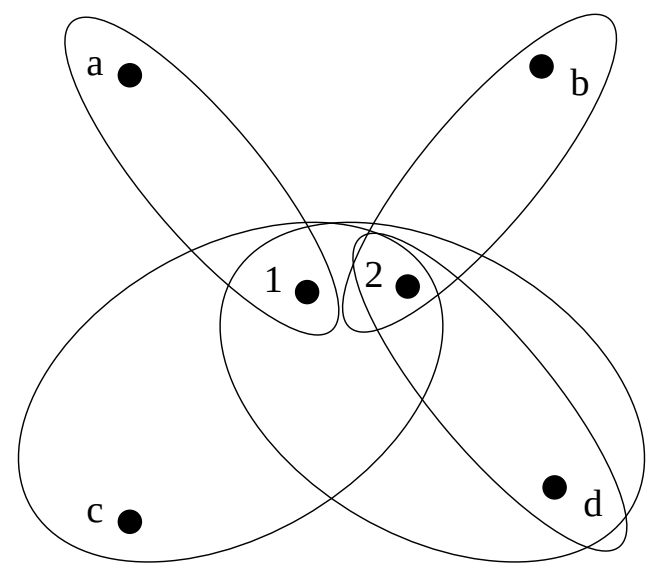

Figure 5: 6-qubit hypergraph whose state has a 1-dimensional stabilizer generated by a weight 4 stabilizing element.

among products of operators $C_e$ on $m$-qubit state space, with real coefficients $c_S$ (here we set $\prod_{e\in\emptyset} C_e = \mathbb{1}$). Let $G = (V, E)$ be the hypergraph given by the following collections of vertices and hyperedges. The vertex set $V$ consists of vertices labeled 1 through $m$, together with a new vertex $v_S$ for each subset $S \subseteq \mathcal{P}(\{1, \ldots, m\})$ for which the coefficient $c_S$ in (23) is nonzero. To construct the hyperedges: for each $S$ such that $c_S \neq 0$, and for each element $e \in S$, form the union $v_S \cup e$. To summarize, we have the following.

$$\begin{aligned} V &= \{1, 2, \ldots, m\} \cup \{v_S \colon S \subseteq \mathcal{P}(\{1, \ldots, m\}), c_S \neq 0\} \\ E &= \{v_S \cup e \colon e \in S, S \subseteq \mathcal{P}(\{1, \ldots, m\})\} \end{aligned}$$

By this construction, we have arranged that $P_{v_S} = \prod_{e\in S} C_e$, so that for

$$M = \sum_S c_S X_S$$

we have $\Phi_M$ equal to the left side of (23). Thus $iM$ is a nontrivial element of $K_{\psi_G}$.

Here is a summary statement of the preceding discussion.

**(5.7) Proposition.** *Suppose that there is a nontrivial linear relation among products of $C_e$ operators acting on a set of qubits. Then by adding qubits and choosing hyperedges in the augmented set of vertices, there is a hypergraph state with a nontrivial local unitary algebra stabilizing element with the same number of nonzero terms as in the linear relation among the $C_e$'s.*

The construction summarized in (5.7) is *not* a one-to-one correspondence between the set of linear relations among products of generalized controlled-$Z$ gates and local unitary algebra elements. For example, consider the 4-qubit hypergraph state with vertices $1, 2, 3, 4$ and hyperedges $123, 124, 14, 24$ (see Figure 6. The operators $Z_1, Z_2, X_3, X_4$ act by $C_1, C_2, C_{1,2}, C_1C_2C_{1,2}$, respectively, so for $M = -Z_1 - Z_2 + X_3 + X_4$ we have $\Phi_M$ equal to the left side of (21), and therefore $iM$ is in the local unitary stabilizer algebra of this state. Thus the relation (21) corresponds to (different) weight 4 stabilizing elements for distinct states, one with 4 qubits and one with 6 qubits.

Next we show how to use (5.7) to construct hypergraph states with arbitrarily high weight LU stabilizing elements, such that there are no lower weight stabilizing elements. Once again we start with an example, and then generalize. One can check that the following relation holds.

$$C_{234}C_{341} - C_{341}C_{412} + C_{412}C_{123} - C_{123}C_{234} = 0 \tag{24}$$

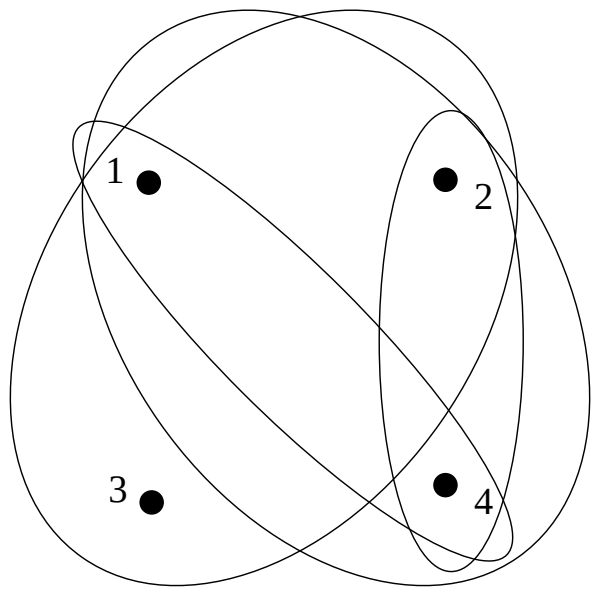

Figure 6: 4-qubit hypergraph whose state has a 1-dimensional stabilizer generated by a weight 4 stabilizing element.

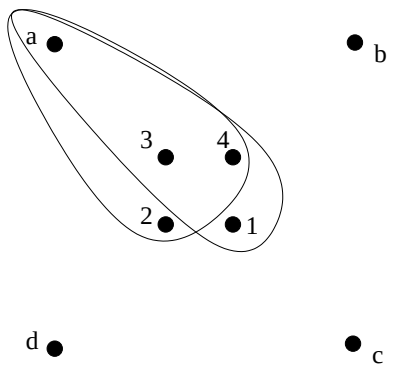

Figure 7: 8-qubit hypergraph whose state has a 1-dimensional stabilizer generated by a weight 4 stabilizing element. Only the hyperedges containing qubit $a$ are shown. Qubits $b, c, d$ each have two hyperedges that are 90 degree rotations of those at $a$.

Applying the construction (5.7) yields an 8-qubit hypergraph state (see Figure 7) with vertices $a, b, c, d, 1, 2, 3, 4$ and hyperedges

$$a234, a341, b341, b412, c412, c123, d123, d234$$

so that the resulting hypergraph state is stabilized by the LU algebra element

$$X_a - X_b + X_c - X_4.$$

One can check (in principle by hand, but more easily with a computer algebra system) that $K_{\psi_G}$ is 1-dimensional, so we conclude that $K_{\psi_G}$ has no generators of weight less than 4.

The relation (24) is the specific case $r = 2$ of the following (the proof is in the appendix).

**(5.8) Proposition.** *Let $r \geq 2$ be a positive integer and let $S$ denote the set $\{1, 2 \ldots, 2r\}$. Then we have the following relation among products of $2r$-qubit controlled-Z gates.*

$$\sum_{j=1}^{r} \left( C_{S\setminus\{2j-1\}} C_{S\setminus\{2j\}} - C_{S\setminus\{2j\}} C_{S\setminus\{2j\oplus 1\}} \right) = 0 \tag{25}$$

*where '$\oplus$' denotes addition mod $2r$ (so that $2r \oplus 1 = 1$).*

Applying the construction (5.7) to (25), yields a $4r$-qubit hypergraph state with vertices

$$1, 2, \ldots, 2r, a_1, b_1, a_2, b_2, \ldots, a_r, b_r \tag{26}$$

and hyperedges

$$a_j \cup (S \setminus \{2j-1\}), a_j \cup (S \setminus \{2j\}), b_j \cup (S \setminus \{2j\}), b_j \cup (S \setminus \{2j \oplus 1\}) \tag{27}$$

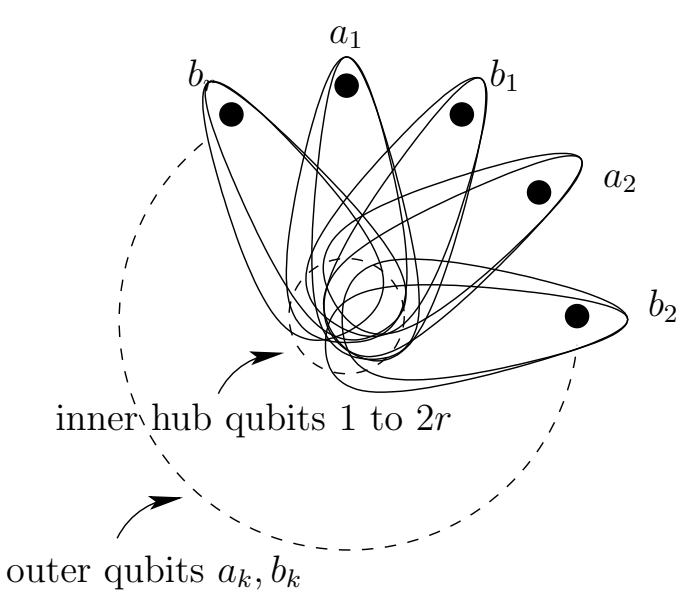

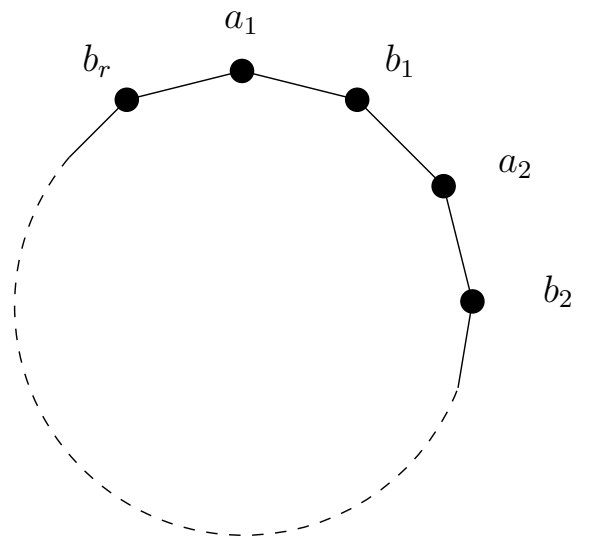

Figure 8: 4r-qubit hypergraph (left) whose state has a 1-dimensional stabilizer generated by a weight $2r$ stabilizing element. Each outer qubit $a_k, b_k$ is contained in 2 hyperedges, generalizing the detail in Figure 7. The corresponding essential hypergraph (right) is the $2r$-gon.

for $1 \leq j \leq r$, so that the resulting hypergraph state is stabilized by the weight $2r$ LU algebra element

$$(X_{a_1} - X_{b_1}) + (X_{a_2} - X_{b_2}) + \cdots + (X_{a_r} - X_{b_r}). \tag{28}$$

The essential hypergraph $\hat{G}$ for this state is a 2-uniform $2r$-gon where essential hyperedges connect adjacent pairs in the list $a_1, b_1, a_2, b_2, \ldots, a_r, b_r$. See Figure 8.

It remains to be shown that the weight of this stabilizing element (28) is *minimal*, i.e., there is no nonzero stabilizing element of lower weight. In fact, we show that the LU stabilizer algebra is 1-dimensional, so that all stabilizing algebra elements are proportional to $\sum_j (X_{a_j} - X_{b_j})$. We state this fact here, and the proof is in the appendix.

**(5.9) Proposition.** *Let $r \geq 2$ be a positive integer, and let $G$ be the hypergraph described above with $4r$ vertices (26) and $4r$ hyperedges (27) with corresponding hypergraph state $|\psi_G\rangle$ and LU stabilizer algebra $K_{\psi_G}$. Then $K_{\psi_G}$ is 1-dimensional, with basis element $\sum_{j=1}^{r}(X_{a_j} - X_{b_j})$.*

This establishes that there is no upper bound to the weights of generators of LU stabilizers of hypergraph states. It also establishes the existence of an infinite family of hypergraph states that are LU-inequivalent to graph states (LU stabilizers of graph states are generated by weight 2 elements).

# 6 Symmetric hypergraph states

While the LU stabilizer algebra $K_\psi$ recovers a great deal of the structure of the LU symmetry group $\mathrm{Stab}_\psi$, the exponential map $M \to \exp(M)$ maps all of $K_\psi$ surjectively onto the connected component of $\mathrm{Stab}_\psi$ that contains the identity. In general, $\mathrm{Stab}_\psi$ is a (semidirect) product

$$\mathrm{Stab}_\psi = \exp(K_\psi) \times H$$

where $H$ is a finite group (see, for example [19]).

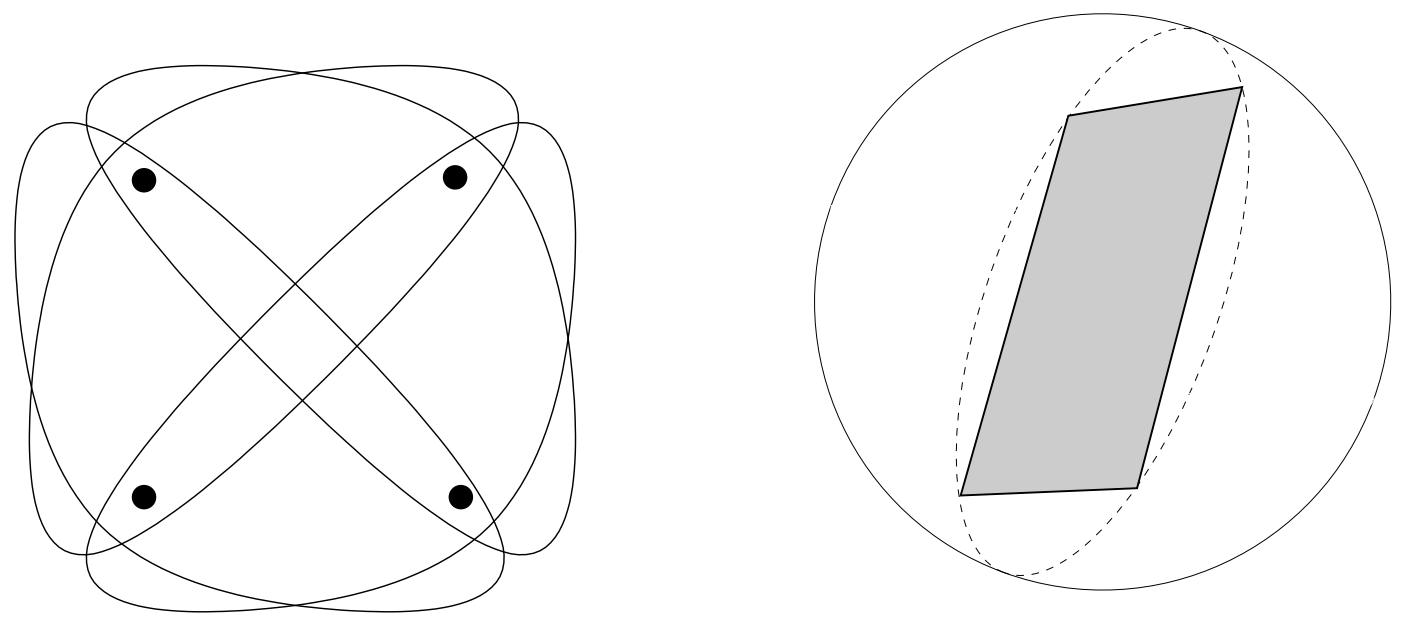

Figure 9: The 4-qubit 3-complete hypergraph (left) and its Majorana points on the Bloch sphere (right).

There is good reason to be interested in the discrete group $H$. The theory and applications of graph states rest on the discrete Pauli stabilizer group, which is a finite subgroup of size $2^n$ of the LU stabilizer of an $n$-qubit graph state $|\psi\rangle$, for which $K_\psi$ may be trivial.

In general, finding $H$ is difficult, but in the special case of permutationally invariant states (that is, invariant under any permutation of the qubits), also called symmetric states, affords additional structure that makes it possible to find discrete symmetries. We begin with an example. Consider the 4-qubit hypergraph on vertices $1, 2, 3, 4$ with hyperedges $123, 124, 134, 234$. Because the hypergraph is 3-uniform and has all possible hyperedges of size 3, it is clear that the state is permutationally invariant. This state, which we call the *3-complete* 4-qubit hypergraph state, has the following properties (the proof is in the appendix). See Figure 9.

**(6.10) The 4-qubit 3-complete hypergraph state.** *The 4-qubit 3-complete state is the symmetrization*

$$\alpha \sum_{\pi \in S_4} |\psi_{\pi(1)}\rangle\, |\psi_{\pi(2)}\rangle\, |\psi_{\pi(3)}\rangle\, |\psi_{\pi(4)}\rangle$$

*of the product of four 1-qubit states $|\psi_1\rangle, |\psi_2\rangle, |\psi_3\rangle, |\psi_4\rangle$ that lie on the corners of a rectangle inscribed on a great circle of the Bloch sphere. Here, $S_4$ denotes the group of permutations of the vertices, and $\alpha$ is a normalization factor. The 4-qubit 3-complete state has a 4-element discrete symmetry group corresponding to the symmetries of the rectangle: the identity, $Y^{\otimes 4}$, $(aX + bZ)^{\otimes 4}$, and $(-bZ + aX)^{\otimes 4}$. The three nontrivial symmetries correspond to the Bloch sphere rotations that are half-turns, respectively, about the $Y$ axis, an axis in the $X, Z$-plane, and another axis in the $X, Z$-plane perpendicular to the first.*

Two important features of this example hold in general. First, it is clear that if we choose integers $m_1, m_2, \ldots, m_r$ in the range $1 \le m_k \le n$, and choose the hyperedges of an $n$-vertex hypergraph to be all possible subsets of sizes $m_1, m_2, \ldots, m_r$, then the resulting hypergraph state is permutationally invariant. Conversely, if a permutationally invariant hypergraph state has a hyperedge of size $m$, then all subsets of the vertex set with size $m$ must also be hyperedges. We record this observation.

**(6.11) Proposition/Definition.** *An $n$-qubit hypergraph state $|\psi\rangle$ is permutationally invariant if and only if there are integers $m_1, m_2, \ldots, m_r$ in the range* $1 \leq m_k \leq n$ *such that the hyperedges of* $|\psi\rangle$ *are all possible subsets of the vertex set of sizes* $m_1, m_2, \ldots, m_r$. *We call such a state the* $(m_1, m_2, \ldots, m_r)$*-complete* $n$*-qubit hypergraph state.*

Second, it is clear that for any choice of $n$ 1-qubit states $|\psi_1\rangle, \ldots, |\psi_n\rangle$, the symmetrized product

$$\alpha \sum_{\pi \in S_4} \left|\psi_{\pi(1)}\right\rangle \left|\psi_{\pi(2)}\right\rangle \cdots \left|\psi_{\pi(n)}\right\rangle \tag{29}$$

(where $\alpha$ is a normalization factor) is permutationally invariant. Not obvious, but true nonetheless, is that the converse holds (see, for example, Bastin et al. [20]).

**(6.12) Proposition.** *Let* $|\psi\rangle$ *be an* $n$*-qubit permutationally invariant state. Then there exist* $n$ *1-qubit states (not necessarily distinct) such that* $|\psi\rangle$ *is their symmetrized product as in (29) above.*

The 1-qubit states whose symmetrized product is $|\psi\rangle$ are called the *Majorana points* for $|\psi\rangle$.

The correspondence between symmetric states and Majorana points is the following. Given a symmetric state $|\psi\rangle = \sum_I |I\rangle$ expanded in the computational basis, permutation invariance implies that if $\mathrm{wt}(I) = \mathrm{wt}(J)$, then $c_I = c_J$ ($\mathrm{wt}(I)$ denotes the Hamming weight, or number of 1's, in the bit string $I$). Gathering terms $|I\rangle$ by weight, we may write $|\psi\rangle$ as a sum

$$|\psi\rangle = \sum_{k=0}^{n} d_k \left|D_n^{(k)}\right\rangle \tag{30}$$

where $\left|D_n^{(k)}\right\rangle$ denotes the weight $k$ Dicke state

$$\left|D_n^{(k)}\right\rangle = \frac{1}{\sqrt{\binom{n}{k}}} \sum_{I:\ \mathrm{wt}(I)=k} |I\rangle\,. \tag{31}$$

Let $\lambda_1, \ldots, \lambda_n$ be the roots (not necessarily distinct) of the polynomial

$$p(z) = \sum_{k=0}^{n} (-1)^k \sqrt{\binom{n}{k}} d_k z^k \tag{32}$$

(called the Majorana polynomial for $|\psi\rangle$). The Majorana points for $|\psi\rangle$ are the inverse stereographic projections of the conjugate roots $\lambda_k^*$.

Here is an immediate consequence of the Majorana construction.

**(6.13) Observation.** *Among all symmetric states, symmetric hypergraph states are characterized by either of the following properties.*

*(i) The coefficient* $d_k$ *in the Dicke basis expansion (30) satisfies* $\sqrt{2^n} d_k = \pm\sqrt{\binom{n}{k}}$.

*(ii) The coefficient of* $z^k$ *in* $\sqrt{2^n} p(z)$*, where* $p(z)$ *is the Majorana polynomial (32), are* $\pm\binom{n}{k}$.

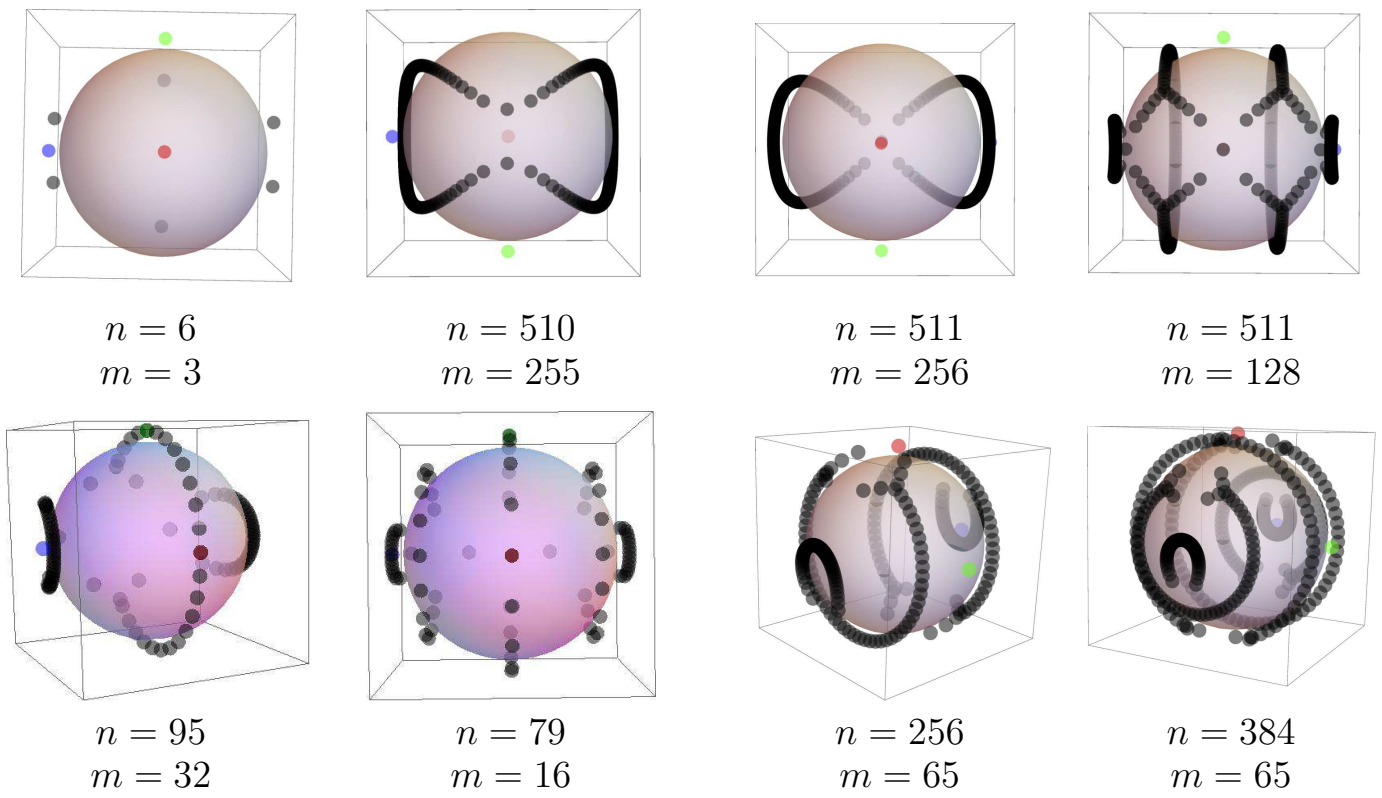

Figure 10: Majorana configurations with half-rotation about $X$- and $Y$-axes (about the red and green dots, respectively, in the color version) symmetry, corresponding to some $n$-qubit $m$-complete hypergraph states with $X^{\otimes n}$ or $Y^{\otimes n}$ symmetry.

LU symmetries for the symmetric state $|\psi\rangle$ are captured in the Majorana point configuration picture as follows (see [21] Theorem 2).

**(6.14) Proposition.** *Let $|\psi\rangle$ be a symmetric state such that $K_\psi = 0$. The local unitary group element $g = (e^{i\theta}, g_1, g_2, \ldots, g_n)$ stabilizes $|\psi\rangle$ if and only if $g_1$, acting as a rotation of the Bloch sphere, permutes the set of Majorana points. Moreover, we have*

$$e^{i\theta} \otimes g_1 \otimes \cdots \otimes g_n = (g_1)^{\otimes n}.$$

Comment: Because an $n$-qubit $m$-complete state has no shared reduced sets, results (5.4) and (6.11) imply that the hypothesis $K_\psi = 0$ is satisfied for any symmetric hypergraph state with the smallest $m \geq 3$.

Figure 10 shows computer generated plots of Majorana configurations for some symmetric hypergraph states. Notable in our searches so far is that all the symmetries we have found are order 2, that is, half rotations about some axis of the Bloch sphere. Also interesting is that this half rotation is the only nontrivial element of all the LU symmetry groups we have found so far, with the exception of the 4-qubit 3-complete state that has a 4-element symmetry group.

A computer plot of Majorana points can suggest possible symmetries, but is not a proof. At present, we do not have a general method for proving the existence or nonexistence of discrete LU symmetries for symmetric hypergraph states. However, we do have proof methods for some specific families stabilized by $X^{\otimes n}$ and $Y^{\otimes n}$. Here is a statement. [Note: The proof in the appendix amounts to the study of the palindromic (or *antipalindromic*) properties of the state vector. This reduces, in turn, to a study of patterns in the mod 2 Pascal's triangle. A delightful 1852 result of Kummer, giving a criterion that determines evenness or oddness of binomial coefficients, provides the key analytic tool.]

**(6.15) Proposition.**

*(a) Let $j \geq 1$, $\ell \geq 0$, let $m$ be in the range $0 \leq m \leq 2^j - 1$ and let $n = (\ell + 1)2^j + m - 1$. The $n$-qubit $m$-complete hypergraph state is stabilized*

*by $X^{\otimes n}$.*

(b) *Let $j \geq 1$, $\ell \geq 1$ be odd, let $m = 2^j$ and let $n = (\ell+1)2^j + m - 1$. Then the computational basis state vector for the $n$-qubit $m$-complete hypergraph state is stabilized by $X^{\otimes n}$.*

(c) *Let $j \geq 1$, let $\ell \geq 0$ be even, let $m = 2^j$ and let $n = (\ell+1)2^j + m - 1$. Then the computational basis state vector for the $n$-qubit $m$-complete hypergraph state is stabilized by $-X^{\otimes n}$.*

(d) *For $j \geq 1$, $\ell \geq 1$, let $n = 2^{j+1}\ell$ and let $m = 2^j + 1$. The $n$-qubit $m$-complete hypergraph state is stabilized by $Y^{\otimes n}$.*

It is interesting to note that the family in part (c) of the above proposition, which we discovered by searching for discrete symmetries in symmetric hypergraph states, is precisely the same as the family in equation (53) in [8] used to construct contextuality inequalities.

# 7 Reduced Density Matrices

In this section we study the partial trace operation on hypergraph states and use the results to make a statement about reconstructing an $n$-qubit hypergraph state from one of its $(n-1)$ qubit reduced density matrices.

Given an $n$-vertex hypergraph $G = (V, E)$ and a choice of a vertex $a \in V$, there are two natural ways to construct an $(n-1)$-vertex hypergraph with vertex set $V \setminus \{a\}$. One is to delete all the hyperedges that contain $a$ and keep all the others. Another is to shrink all the hyperedges by simply omitting $a$ from all the hyperedges in which it appears (in the case that there are hyperedges $f, g$ such that $f = g \setminus \{a\}$, we discard both $f, g \setminus \{a\}$ because $C_f = C_{g\setminus\{a\}}$ so $C_f C_{g\setminus\{a\}} = \mathbb{1}$). See Figure 11 for examples. The delightful trace result is that tracing the density matrix $\rho_G$ over the qubit $a$ is the equal mixture of the two hypergraph states formed by deleting and shrinking. Here is the formal statement (the proof is in the appendix). We write $\rho_G$ to denote the density matrix $|\psi_G\rangle\langle\psi_G|$.

**(7.16) Proposition (partial trace over one qubit).** *Let $G = (V, E)$ be an $n$-qubit hypergraph state. Define $(n-1)$-vertex hypergraphs $D_a G$ and $S_a G$ (D for 'delete' and S for 'shrink') by*

$$\begin{aligned} D_a G &= (V \setminus \{a\}, \{e \in E \colon a \notin e\}) \\ S_a G &= (V \setminus \{a\}, \{e \in E \colon a \notin e\} \bigtriangleup \{e \setminus \{a\} \colon a \in e \in E\}) \end{aligned}$$

*(where $A \bigtriangleup B$ denotes the symmetric difference $A \setminus B \cup B \setminus A$ of sets $A, B$). Then we have*

$$\operatorname{tr}_a \rho_G = \frac{1}{2}\left(\rho_{D_a G} + \rho_{S_a G}\right). \tag{33}$$

*In particular, the partial trace over any one qubit of a hypergraph state is a mixture of hypergraph states. It follows that the partial trace over any subsystem of a hypergraph state is a mixture of hypergraph states.*

We can think of the operators $D_a, S_a$ as operators on the space of density matrices of hypergraph states, so that the trace formula (33) can be written

$$\operatorname{tr}_a = \frac{1}{2}(D_1 + S_1). \tag{34}$$

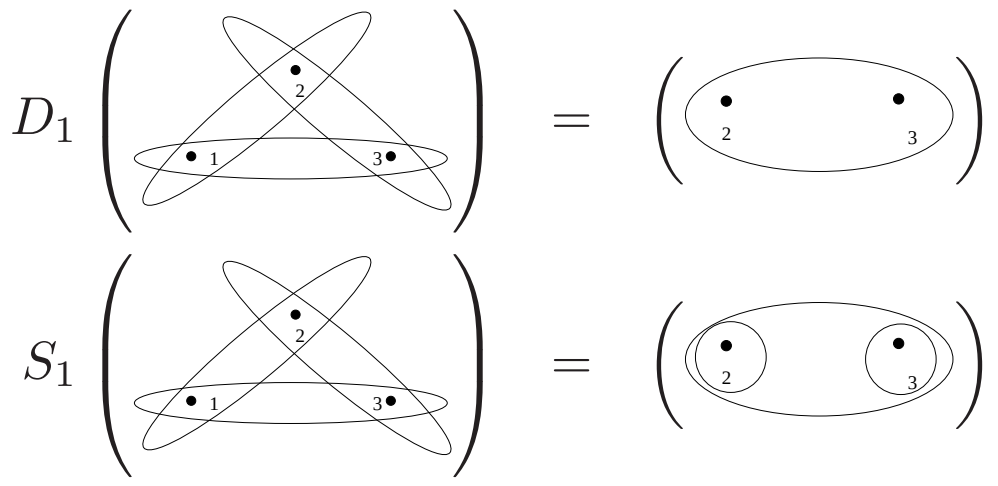

Figure 11: Examples of delete and shrink operators

It is clear that $D_i D_j = D_j D_i$ for $i \neq j$, and likewise, it is clear that the pairs $D_i, S_j$ and $S_i, S_j$ commute for $i \neq j$. Thus, it makes sense to write $D_{ij}$ to denote the product $D_i D_j$, and so on. In general, given a subset $U$ of the set of qubits $\{1, 2, \dots, n\}$, we write

$$D_U = \prod_{a \in U} D_a \tag{35}$$

$$S_U = \prod_{a \in U} S_a. \tag{36}$$

Using this notation, we can write the partial trace over two qubits, say qubits $1, 2$, as follows.

$$\mathrm{tr}_{1,2} = \frac{1}{4}(D_1 + S_1)(D_2 + S_2) \tag{37}$$

$$= \frac{1}{4}(D_1 D_2 + D_1 S_2 + S_1 D_2 + S_1 S_2) \tag{38}$$

$$= \frac{1}{4}(D_{12} + D_1 S_2 + D_2 S_1 + S_{12}) \tag{39}$$

In general, we have the following expression for partial trace over a subsystem $U \subseteq \{1, 2, \dots, n\}$.

$$\mathrm{tr}_U = \frac{1}{2^{|U|}} \sum_{T \subseteq U} D_T S_{T^c} \tag{40}$$

See Figure 12 for examples, shown using a diagram shorthand where the picture of a hypergraph denotes the density matrix of the corresponding hypergraph state.

As an application, we consider the question of reconstructing an unknown hypergraph state from its $(n-1)$-party reduced states. That is, if we are given $\mathrm{tr}_a \, \rho_G$ can we say what $G$ is? First, we answer the question of how different two mixtures of hypergraph states can be, when each is a mixture of two hypergraph states. The proof is in the appendix.

**(7.17) Proposition.** *Let $H, K, H', K'$ be hypergraphs, and suppose that*

$$\rho_H + \rho_K = \rho_{H'} + \rho_{K'}.$$

*Then either*

(a) $|\psi_H\rangle = |\psi_{H'}\rangle$ *and* $|\psi_K\rangle = |\psi_{K'}\rangle$*, or*

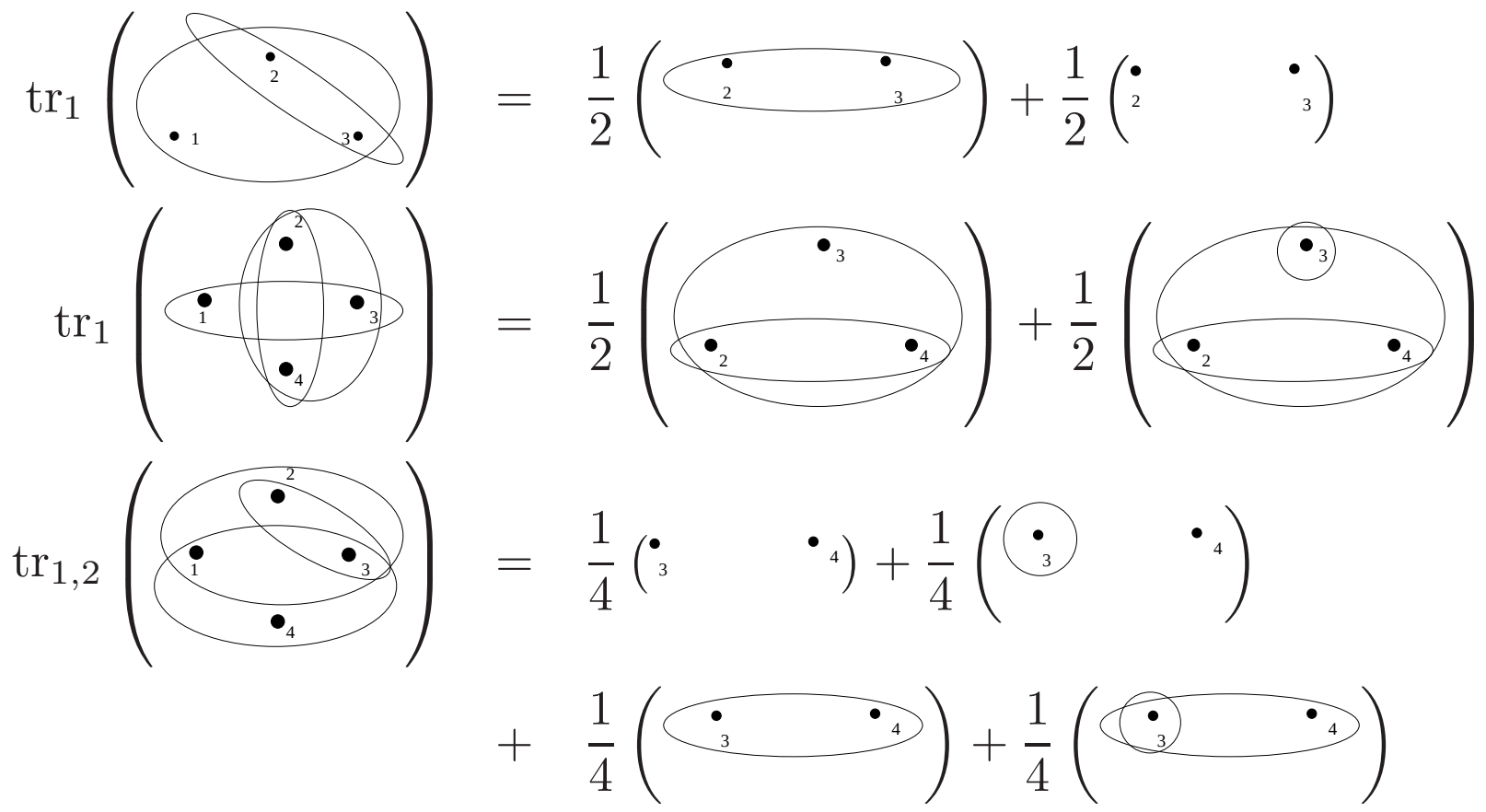

Figure 12: Partial trace “diagram calculus” examples

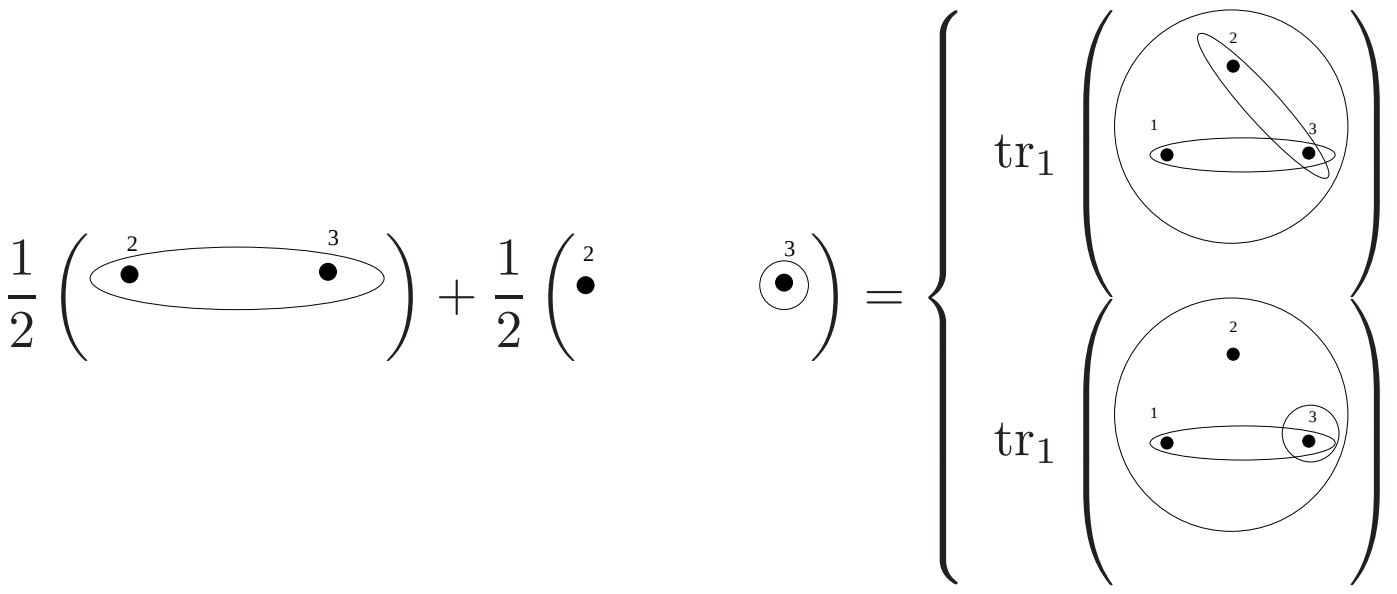

Figure 13: Example: possible hypergraphs sharing the same $(n-1)$-qubit reduced density matrices

*(b)* $|\psi_H\rangle = |\psi_{K'}\rangle$ *and* $|\psi_K\rangle = |\psi_{H'}\rangle$.

Here is the result on how an $n$-qubit hypergraph state is determined by its $(n-1)$-qubit reduced density matrices (the proof is in the appendix).

**(7.18) Proposition.** *Suppose that* $\operatorname{tr}_1 \rho_G = \frac{1}{2}(\rho_H + \rho_K)$*. Then* $|\psi_G\rangle$ *is either*

1. $\frac{1}{\sqrt{2}}\left(|0\rangle|\psi_H\rangle \pm |1\rangle|\psi_K\rangle\right)$, *or*

2. $\frac{1}{\sqrt{2}}\left(\pm|1\rangle|\psi_H\rangle + |0\rangle|\psi_K\rangle\right)$

*where the '$\pm$' sign is* $-1$ *if* $\{1\}$ *is a hyperedge in* $G$ *and is* $+1$ *otherwise.*

Figure 13 illustrates the two possibilities in the case that $\{1\}$ is not a hyperedge in $G$.

# 8 Conclusion and outlook

We have calculated the action of an arbitrary local unitary algebra element on a hypergraph state (Proposition (4.1)), and have used that calculation to describe the structure of local unitary algebra symmetries of $m$-uniform hypergraph states with $m \geq 3$ (Propositions (5.2)–(5.4)). Using relations in the algebra of generalized controlled-$Z$ gates, we have constructed families of states with high dimensional symmetry groups and with 1-dimensional symmetry groups generated by elements of high weight (Propositions (5.5)–(5.9)). We have also presented families of symmetric hypergraph states that have nontrivial discrete symmetries (Proposition (6.15)). Finally, we have characterized the set of $n$-qubit hypergraph states that can share the same $(n-1)$-party reduced states (Proposition (7.18)). Collectively, these results provide an array of tools and techniques that lead to a series of observations that, admittedly short of a full classification of hypergraph state multipartite entanglement types, shed new insights and lead naturally to further development. Some natural follow-up investigations include the following.

- Find a nice description of the algebra of generalized controlled-$Z$ gates. For example, characterize it by giving generators and relations.
- All discrete symmetries of permutationally invariant states that we have found have order 2. Is this true in general?
- The only permutationally invariant state that we have found whose local unitary symmetry group has more than two elements is the 4-qubit 3-complete state. Is this really the only possibility?
- To what extent do reduced density matrices determine a hypergraph state? How far can our result on $(n-1)$-party subsystems be extended?
- What is the connection between the families of states we identify with $X^{\otimes n}$ symmetry and the contextuality results of Gühne et al. [8]?

We expect that further investigations of hypergraph states will continue to bear interesting results.

**Acknowledgments.** DWL thanks Barbara Kraus, Martí Cuquet, Marcus Huber, and Ottfried Gühne for stimulating discussions. This work is supported by National Science Foundation award PHY-1211594 and a Lebanon Valley College research grant.

# A Proofs of Propositions

### Proof of (4.1)

We start with an elementary observation that

$$\wedge_e I + \wedge_{e \setminus a} I = \wedge_e I_a \pmod 2 \tag{41}$$

for all qubits $a$, hyperedges $e$, and $n$-bit strings $I = i_1 i_2 \cdots i_n$, and where $I_a$ denotes the $n$-bit string $i_1 i_2 \cdots i_a^c \cdots i_n$ obtained from $I$ by flipping the bit in position $a$. Indeed, if $i_a = 0$, then $\wedge_{e\setminus a} I = \wedge_e I_a$ and $\wedge_e I = 0$, and if $i_a = 1$ then $\wedge_e I = \wedge_{e\setminus a} I$ and $\wedge_e I_a = 0$.

Now we derive, starting with ($\sqrt{2^n}$ times) the left side of (12).

$$\begin{aligned}
\sqrt{2^n} X_a |\psi_G\rangle &= X_a \sum_I (-1)^{\sum_{e\in E} \wedge_e I} |I\rangle \qquad \text{(use (7))} && (42)\\
&= \sum_I (-1)^{\sum_{e\in E} \wedge_e I} |I_a\rangle \qquad \text{(use } X_a |I\rangle = X |I_a\rangle\text{)} && (43)\\
&= \sum_I (-1)^{\sum_{e\in E} \wedge_e I_a} |I\rangle && (44)\\
&= \sum_I (-1)^{\sum_{e\in E} (\wedge_{e\setminus a} I + \wedge_e I)} |I\rangle \qquad \text{(observation (41))} && (45)\\
&= \sqrt{2^n} \left( \prod_{e:\, a\in e} C_{e\setminus a} \right) \left( \prod_e C_e \right) |+\rangle^{\otimes n} \qquad \text{(use (6))} && (46)\\
&= \sqrt{2^n} \left( \prod_{e:\, a\in e} C_{e\setminus a} \right) |\psi_G\rangle \qquad \text{(use (3) and (7) again)} && (47)
\end{aligned}$$

This establishes (12). Equation (13) follows immediately from the relation $ZX = iY$.

**Proof of (5.3)**

Applying both sides of (19) to the computational basis vector $|0\rangle^{\otimes n}$, we get

$$\theta + \sum_a (r_a + t_a) = 0. \tag{48}$$

Applying both sides of (19) to $|1_b\rangle$, we get

$$\theta + \sum_a r_a - t_b + \sum_{a\neq b} t_a = 0 \tag{49}$$

(here we use the assumption $k \geq 3$ to guarantee that $P_a$ acts on 2 or more qubits, so that we are certain there are no minus signs on any $r_a$'s in (49)). Adding (A) plus (49), we conclude that $t_b = 0$. This holds for all qubits $b$, so all the coefficients $t_a$ are zero. A similar argument, applying both sides of (20) to $|0\rangle^{\otimes n}$ and $|1_b\rangle$, shows that all the coefficients $s_a$ are zero.

**Proof of (5.4)**

Proof. Let $f$ be a hyperedge containing vertex $b$ whose reduced set $f \setminus \{b\}$ is not equal to $g \setminus \{c\}$ for all qubits $c$ and hyperedges $g$ containing $c$. Consider $M = (\theta + \sum_k r_k X_k)$ such that $iM \in K_{\psi_G}$ (Theorem (5.3) guarantees that any stabilizing element is of this form). Applying $M$ to the computational basis vector $|0\rangle^{\otimes n}$, we get

$$\theta + \sum_a r_a = 0$$

and applying $M$ to the computational basis vector $|1_{f\setminus b}\rangle$ yields

$$\theta - r_b + \sum_{a\neq b} r_a = 0$$

(by assumption, $f \setminus b$ is not the reduced set for any qubit except $b$, and by assumption of $m$-uniformity, there is no subset of $f \setminus b$ that is a reduced set for any qubit, so $r_b$ is the only one of the $r$ coefficients with a minus sign). Adding these two equations, we get $r_b = 0$. □

**Proof of (5.5)**

By the definition of essential hypergraph, we have $P_1 = P_2 = \cdots P_{|\hat{V}|}$, so it is clear that $X_1 - X_j$, which acts by $P_1 - P_j$, is a stabilizing element for qubit labels $1, 2 \ldots, |\hat{V}|$ of the essential vertices. It remains to be shown that the $i(X_1 - X_j)$ form a basis for the entire stabilizer algebra. By theorem (5.3), an arbitrary stabilizer algebra element is of the form $i(\theta + \sum_a r_a X_a)$, where we know that the index $a$ runs over only the essential vertex set by Proposition (5.4). Thus, we can take that linear combination

$$M = \theta + \sum_a r_a X_a + \sum_{b \neq a} r_b (X_1 - X_b) = \theta + \left( \sum_a r_a \right) X_1$$

so that $iM \in K_{\psi_G}$. By (18) and (19), we have $\theta + (\sum_a r_a) P_1 = 0$, so we conclude that $\theta = \sum_a r_a = 0$. This shows that the elements $i(X_1 - X_j)$ span $K_{\psi_G}$. This completes the proof of the proposition.

**Proof of (5.6)**

By (5.3) and (5.4), any element $M$ with $iM$ in $K_{\psi_G}$ is of the form $M = \theta + \sum_a r_a X_a$, where the index $a$ ranges over the essential vertices. Choose 3 essential vertices $u, v, w$ connected by adjacent essential hyperedges $uv, vw$ (the hypotheses guarantee that this is possible). Let $S, T$ be the reduced sets corresponding to essential hyperedges $uv, vw$, respectively. Applying stabilizing element $M$ to $|0\rangle^{\otimes n}$, $|1_S\rangle$, $|1_T\rangle$, $|1_{S \cup T}\rangle$, we get the following equations (the hypothesis that shared reduced sets are distinct for different pairs of vertices guarantees that there are exactly two minus signs in the last three equations).

$$\sum_a r_a = 0 \quad (50)$$

$$-r_u - r_v + \sum_{a \neq u,v} r_a = 0 \quad (51)$$

$$-r_v - r_w + \sum_{a \neq v,w} r_a = 0 \quad (52)$$

$$-r_u - r_w + \sum_{a \neq u,w} r_a = 0 \quad (53)$$

Summing equations, we have $r_u = r_w = -r_w$, so $r_u = 0$. The same holds for all essential vertices.

**Proof of (5.8)**

Let $A_j = C_{S \setminus \{2j-1\}} C_{S \setminus \{2j\}}$ and $B_j = C_{S \setminus \{2j\}} C_{S \setminus \{2j \oplus 1\}}$ for $1 \leq j \leq r$, where $\oplus$ denotes addition mod $2r$, so that the left side of (25) is $\sum_{j-1}^{r} (A_j - B_j)$. For $T \subseteq S$ of size $|T| \neq 2r - 1$, we have

$$A_j |1_T\rangle = B_j |1_T\rangle = |1_T\rangle$$

so $\sum_j (A_j - B_j) |1_T\rangle = 0$. For $|T| = 2r - 1$, say $T = S \setminus \{k\}$ for some $1 \leq k \leq 2r$, we consider two cases. For even $k = 2\ell$, we have $A_j |1_T\rangle = B_j |1_T\rangle = |1_T\rangle$ for $j \neq \ell$ and we have $A_\ell |1_T\rangle = B_\ell |1_T\rangle = -|1_T\rangle$, so that $\sum_j (A_j - B_j)|1_T\rangle = 0$. For odd $k = 2\ell - 1$, we have $A_j |1_T\rangle = B_{j-1}|1_T\rangle = |1_T\rangle$ for $j \neq \ell$ and we have $A_\ell |1_T\rangle = B_{\ell-1}|1_T\rangle = -|1_T\rangle$ (where the value of the subscript $\ell - 1$ is understood to be taken mod $2r$), so again we have cancellation in pairs $\sum_j (A_j - B_j)|1_T\rangle = 0$. Thus we see that (25) holds in all cases.

**Proof of (5.9)**

By (5.3), we may assume that an element of the LU stabilizer algebra $K_{\psi_G}$ has the form $iM$ for some $M$ of the form

$$M = \theta + \sum_{j=1}^{r} (\alpha_j X_{a_j} + \beta_j X_{b_j}) + \sum_{k=1}^{2r} \gamma_k X_k$$

for some real $\theta, \alpha_j, \beta_j, \gamma_k$. We begin by showing that the coefficients $\gamma_k$ are all zero. For $1 \leq j \leq r$, let $|u_j\rangle$ and $|v_j\rangle$ denote the standard basis vectors

$$\begin{aligned} |u_j\rangle &= \left|1_{a_j \cup (S \setminus \{2j\})}\right\rangle \\ |v_j\rangle &= \left|1_{a_j \cup (S \setminus \{2j-1\})}\right\rangle \end{aligned}$$

Consider first the case when $k = 2j$ is even. There are only two hyperedges that contain $a_j$, and only one of those contains $2j$, namely the hyperedge $e = a_j \cup (S \setminus \{2j-1\})$. Thus, the operator $C_{e \setminus \{2j\}}$ is the only factor in the product $P_{2j}$ that acts nontrivially on $|u_j\rangle$ (see (8)). Thus we have

$$X_k |u_j\rangle = P_{2j} |u_j\rangle = -|u_j\rangle$$

and for $\ell \neq k$, we have $X_\ell |u_j\rangle = |u_j\rangle$. Thus we have

$$\theta + \sum_j (\alpha_j + \beta_j) + \sum_{\ell \neq k} \gamma_\ell - \gamma_k = 0. \tag{54}$$

Applying $M$ to $|0\rangle^{\otimes 4r}$, we have

$$\theta + \sum_j (\alpha_j + \beta_j) + \sum_k \gamma_k = 0. \tag{55}$$

Adding the two equations above, we get $\gamma_k = 0$. A similar argument using $|v_j\rangle$ in place of $|u_j\rangle$ shows that $\gamma_k = 0$ for $k$ odd.

Next, note that

$$P_{a_j} \left|1_{S\setminus\{2j\}}\right\rangle = P_{b_j} \left|1_{S\setminus\{2j\}}\right\rangle = -\left|1_{S\setminus\{2j\}}\right\rangle \tag{56}$$

and $P_c \left|1_{S\setminus\{2j\}}\right\rangle = \left|1_{S\setminus\{2j\}}\right\rangle$ for $c = a_\ell, b_\ell$ whenever $\ell \neq j$. Adding (56) to (A), we get $\alpha_j + \beta_j = 0$. A similar argument using $\left|1_{S\setminus\{2j-1\}}\right\rangle$ in place of $\left|1_{S\setminus\{2j\}}\right\rangle$, gives $\beta_{j-1} + \alpha_j = 0$. These equations hold for $1 \leq j \leq r$ (where it is understood that addition and subtraction operations in subscripts are mod $2r$), so we have

$$\alpha_1 = -\beta_1 = \alpha_2 = -\beta_2 = \cdots = \alpha_r = -\beta_r.$$

and $\theta = 0$. This concludes the proof.

**Proof of (6.10)**

By (6.12) and (6.14), it suffices to show that the Majorana points for the 4-qubit 3-complete hypergraph state are the four corners of a rectangle on a great circle of the Bloch sphere that passes through the equatorial points $|+\rangle, |-\rangle$, which are the points $\pm i$ in the $X, Y$-plane.

It is straightforward to check that the Majorana polynomial (defined in (32)), is given by

$$p(z) = \frac{1}{4}\left(1 - 4z + 6z^2 + 4z^3 + z^4\right). \tag{57}$$

The roots $\lambda, \lambda^*, \mu, \mu^*$ of

$$p(z) = \frac{1}{4}(\lambda - z)(\lambda^* - z)(\mu - z)(\mu^* - z) \tag{58}$$

(the roots occur in conjugate pairs because the coefficients of $p(z)$ are real) satisfy the following equations.

$$|\lambda|^2|\mu|^2 = 1 \quad \text{(constant term)} \tag{59}$$

$$|\lambda|^2 \operatorname{Re}\mu + |\mu|^2 \operatorname{Re}\lambda = 2 \quad (z \text{ term}) \tag{60}$$

$$\operatorname{Re}(\lambda\mu) + \operatorname{Re}(\lambda\mu^*) + |\lambda|^2 + |\mu|^2 = 6 \quad (z^2 \text{ term}) \tag{61}$$

$$\operatorname{Re}\lambda + \operatorname{Re}\mu + = -2 \quad (z^3 \text{ term}) \tag{62}$$

Write $\lambda = re^{i\theta}, \mu = se^{i\phi}$. The degree 1 coefficient (60) says

$$r^2 s \cos\phi + s^2 r \cos\theta = 2. \tag{63}$$

Factoring out $rs = 1$ (from the degree 0 term (59)) we have

$$r \cos\phi + s \cos\theta = 2. \tag{64}$$

The degree 3 term (62) says

$$r \cos\theta + s \cos\phi = 2. \tag{65}$$

Adding the previous two equations (64) and (65) yields

$$(r + s)(\cos\theta + \cos\phi) = 0 \tag{66}$$

so that we have

$$\cos\theta = -\cos\phi. \tag{67}$$

Thus we have

$$\lambda\mu = rse^{i(\theta+\phi)} = rse^{i\pi} = -1 \tag{68}$$

(again, using $rs = 1$ from the degree 0 term).

Because stereographic projection preserves circles (see [22, 23]), it suffices to show that $\lambda, \mu, i, -i$ lie on a circle together (it is trivial then that $\lambda^*, \mu^*$ also lie on this circle, and the inverse stereographic projection of this circle is a great circle on the Bloch sphere that passes through the points $|+\rangle, |-\rangle$). To show that this is the case, we use a theorem from Möbius geometry that says four points $z_0, z_1, z_2, z_3$ in the complex plane lie together on a euclidean circle or straight line if and only if their cross ratio

$$(z_0, z_1, z_2, z_3) := \frac{z_0 - z_2}{z_0 - z_3}\frac{z_1 - z_3}{z_1 - z_2} \tag{69}$$

is real (see, for example, Henle [24]). Indeed, we have

$$(\lambda, \mu, i, -i) = \frac{\lambda - i}{\lambda + i}\frac{\mu + i}{\lambda - i} \tag{70}$$

$$= \frac{(\lambda\mu + 1) + i(\lambda - \mu)}{(\lambda\mu + 1) - i(\lambda - \mu)} \tag{71}$$

$$= \frac{i(\lambda - \mu)}{-i(\lambda - \mu)} \quad \text{(by (68))} \tag{72}$$

$$\tag{73}$$

which is clearly real. Finally, we note that (67) implies that the inverse stereographic projections of $\lambda, \lambda^*, \mu, \mu^*$ lie on the intersections of two vertical planes that make equal angles with the $X, Z$-plane. This implies that the Majorana points for $|\psi\rangle$ are indeed the four vertices of the claimed rectangle, and the proof is complete.

**Proof of (6.15)**

Let $|\psi_G\rangle$ be the $n$-qubit $m$-complete hypergraph state, so that we have the following expansion in the computational basis.

$$|\psi_G\rangle = \sum_I (-1)^{\binom{\mathrm{wt}\, I}{m}} |I\rangle \tag{74}$$

We begin with the simple observations that $X^{\otimes n}$ takes $|I\rangle$ to $|I^c\rangle$ (where $I^c$ denotes the bitwise complement $i_1^c i_2^c \cdots i_n^c$ of the bit string $I = i_1 i_2 \cdots i_n$) and $Y^{\otimes n}$ takes $|I\rangle$ to $i^n(-1)^{\mathrm{wt}(I)} |I^c\rangle$. Thus $X^{\otimes n}$ and $Y^{\otimes n}$ take computational basis vectors with Hamming weight $w$ to Hamming weight $n - w$, so (74) implies the following necessary and sufficient conditions.

**Observation.** The $n$-qubit $m$-complete hypergraph state

- is stabilized by $X^{\otimes n}$ if and only if

$$\binom{w}{m} = \binom{n-w}{m} \pmod 2 \text{ for all } w,\ 0 \le w \le n, \tag{75}$$

- is stabilized by $-X^{\otimes n}$ if and only if

$$\binom{w}{m} = \binom{n-w}{m} + 1 \pmod 2 \text{ for all } w,\ 0 \le w \le n, \text{ and} \tag{76}$$

- is stabilized by $Y^{\otimes n}$ if and only if $n$ is a multiple of 4 and

$$\binom{w}{m} + w = \binom{n-w}{m} \pmod 2 \text{ for all } w,\ 0 \le w \le n. \tag{77}$$

Another way to state conditions (75) and (76) is to say that the computational state vector $\sum_I c_I |I\rangle$ is a *palindrome* (that is, $c_I = c_{I^c}$ for all $I$) or an *anti-palindrome* (that is, $c_I = -c_{I^c}$ for all $I$), respectively.

In what follows, we make use of base 2 representations of natural numbers. Given a nonnegative integer $d = \sum_{i=0}^r d_i 2^i$, we will write

$$d = d_r d_{r-1} \cdots d_2 d_1 d_0 \tag{78}$$

to denote the base 2 expansion of $d$ as a string of digits $d_i = 0, 1$. We allow "leading zeroes", i.e., it may be that $d_r = 0$.

We will use the following Fact, which is a consequence of an 1852 theorem of Kummer [25].

> **Fact (parity of binomial coefficients).** Let $r, s$ be nonnegative integers with binary expansions $r = \sum_i r_i 2^i$, $s = \sum_i s_i 2^i$. The binomial coefficient $\binom{r+s}{s}$ is odd if and only if there is no index $p$ such that $r_p = s_p = 1$.

Throughout, let $r = w - m$, $r' = n - w - m$, $s = m$ so that

$$\binom{w}{m} = \binom{r+s}{s} \text{ and} \tag{79}$$

$$\binom{n-w}{m} = \binom{r'+s}{s}. \tag{80}$$

Now we proceed with the proof of part (a) of the Proposition. Suppose that $j \geq 1$, $\ell \geq 0$, $0 \leq m \leq 2^j - 1$ and $n = (\ell+1)2^j + m - 1$. We will use the above Fact to establish (75) by proving the following Claim.

**Claim.**

(i) Suppose $w \geq m$ and $n - w \geq m$. Then there is no index $p$ such that $r_p = m_p = 1$ if and only if there is no index $p$ such that $r'_p = m_p = 1$. It then follows from the Fact that $\binom{w}{m}$ is odd if and only if $\binom{n-w}{m}$ is odd.

(ii) Suppose $w < m$ and $n - w \geq m$, so that $\binom{w}{m} = 0$. Then there is no index $p$ such that $r'_p = m_p = 1$, so that $\binom{n-w}{m} = 0 \pmod 2$ by the Fact.

(iii) Suppose $w \geq m$ and $n - w < m$, so that $\binom{n-w}{m} = 0$. Then there is no index $p$ such that $r_p = m_p = 1$, so that $\binom{w}{m} = 0 \pmod 2$ by the Fact.

(iv) Suppose $w < m$ and $n - w < m$, so that $\binom{w}{m} = \binom{n-w}{m} = 0$. Then (75) holds trivially in this case.

Let $\ell = \ell_s \ell_{r-1} \cdots \ell_0$ be the base 2 expansion of $\ell$ (and we allow the possibility that $\ell = 0$ so that $\ell s = \ell_{s-1} = \cdots = \ell_0 = 0$). Then $n - m$ has the form

$$n - m = \ell 2^j + (2^j - 1) \tag{81}$$

$$= \ell_s \cdots \ell_0 \underbrace{11 \cdots 1}_{j \text{ 1's}}. \tag{82}$$

To prove part (i) of the Claim, suppose that there is no index $p$ such $r_p = m_p = 1$, and suppose that $m_{p_0} = 1$ (note that the hypothesis $m \leq 2^j - 1$ implies $p_0 \leq j - 1$). Then $r_{p_0} = 0$. It follows that $w_{p_0} = r_{p_0} + m_{p_0}$ must be 1 (the hypothesis that there is no $p$ such that $r_p = m_p = 1$ means there is no "carrying" in the binary addition of $r$ and $m$). We know that $m - n$ has a 1 in the $p_0$-th position, so the binary digit in the $p_0$-th position of $r' = n - m - w$ is

0. This holds for all $p_0$, so we have established one direction of the if-and-only-if statement made by part (i) of the Claim. Conversely, suppose there is no index $p$ such that $r'_p = m_p = 1$, and suppose that $m_{p_0} = 1$. Then $r'_{p_0} = 0$. It follows that $w_{p_0} = 1$ (because $r' = (n-m) - w$ and $n-m$ has 1 in positions 0 through $j-1$). Since $r = w - m$, it must be that $r_{p_0} = 0$. This concludes the proof of part (i) of the Claim.

Next we consider part (ii) of the Claim. Suppose that $w < m$. Thus we have $w < 2^j - 1$, so that $r' = n - m - w$ has the binary expansion

$$r' = n - m - w = \ell_s \cdots \ell_0 w^c_{j-1} \cdots w^c_0 \tag{83}$$

(where $w^c_k$ denotes the bit complement of the base 2 digit $w_k$). We wish to show that there is an index $p$ such that $r'_p = m_p = 1$. Suppose on the contrary that there is no index $p$ such that $r'_p = m_p = 1$. Then $m_p = 1$ only when $w^c_p = 0$, so only when $w_p = 1$. But this implies $w \geq m$, which contradicts our assumption. We conclude that there is an index $p$ such that $r'_p = m_p = 1$. This establishes part (ii) of Claim 1.

The proof of part (iii) of the Claim is similar to part (ii). Supposing that $n - w < m$, we have $n - w < 2^j - 1$ so we may write the base 2 expansion using at most $j$ nonzero digits

$$n - w = (n-w)_{j-1}(n-w)_{j-2} \cdots (n-w)_0 \tag{84}$$

and thus we have

$$r = w - m = (n-m) - (n-w) = \ell_s \cdots \ell_0 (n-w)^c_{j-1} \cdots (n-w)^c_0. \tag{85}$$

We wish to show that there is an index $p$ such that $r_p = m_p = 1$. Suppose on the contrary that there is no index $p$ such that $r_p = m_p = 1$. Then $m_p = 1$ only when $(n-w)^c_p = 0$, so only when $(n-w)_p = 1$. But this implies $n - w \geq m$, which contradicts our assumption. We conclude that there is an index $p$ such that $r_p = m_p = 1$. This establishes part (iii) of the Claim, and completes the proof of part (a) the Proposition.

The proofs of the remaining parts (b)–(d) of the Proposition use similar logic. Rather than give every detail, here is an outline.

For part (b), using $\ell_0 = 1$, simply note that that arguments in (a) work for $m = 2^j$ by adjoining the $j$th position. For example, expression (82) is now $\ell_s \cdots \ell_1 \underbrace{11\cdots 1}_{(j+1)\ 1\text{'s}}$, the right hand side of (83) is now $\ell_s \cdots \ell_1 w^c_j \cdots w^c_0$, and so on.

For part (c), assume that $m = 2^j$ and $\ell$ is even. Following the same cases of the Claim in the proof of part (a), we need only consider position $p_0 = j$ as the only possible location for which $m_{p_0} = 1$. The case (i) ($w \geq m$ and $n - w \geq m$) cannot happen if $\ell = 0$. If $\ell \geq 2$, it is easy to see that $r'_j = 1$ if and only if $w_j = 1$ if and only if $r_j = 0$. For the case (ii) ($w < m$ and $n - w \geq m$), we have $\binom{w}{m} = 0$ and $r'_j = 0$ so $\binom{n-w}{m} = 1$. Similarly for case (iii) ($w \geq m$ and $n - w < m$), we have $\binom{n-w}{m} = 0$ and $r_j = 0$ so $\binom{w}{m} = 1$. Case (iv) ($w < m$ and $n - w < m$) cannot happen.

For part (d) of the Proposition, we prove the cases of the following modified version of the Claim in the proof of part (a) above.

**Claim 2.**

(i) Suppose that $w$ is even. If $w \geq m$ and $n - w \geq m$, then $r_0 = (w - m)_0 = 1$ and $r'_0 = (n - w - m)_0 = 1$. Thus both sides of (77) are even. Both sides are also even if $w < m$ or $n - w < m$.

(ii) Suppose that $w$ is odd, and suppose $w \geq m$ and $n - w \geq m$. Because $m$ has only two positions with 1's in its binary expansion (namely, $m_j = m_0 = 1$), and because $w$ is odd, the position $p = j$ is the only possible location where $m_p = r_p = r'_p = 1$ could be satisfied. But it turns out that $r_j = (w-m)_j = 0$ if and only if $r'_j = (n - m - w)_j = 1$. It follows that (77) holds. This also holds if $w < m$ or $n - w < m$.

As in part (a) of the Proposition, the proof of part (b) comes down to a careful examination of the binary expansion of $n-m$. From the hypotheses $m = 2^j + 1$, $n = \ell 2^{j+1}$ for some $\ell \geq 1$, $j \geq 1$, we have

$$n - m = d_s \cdots d_1 0 \underbrace{11 \cdots 1}_{j-1 \text{ 1's}} \tag{86}$$

where at least one of the $d_k$ is nonzero. Part (i) of Claim 2 follows easily from this. For part (ii) (the $w$ odd case), assuming $w \geq m$ and $n - w \geq m$, we have

$$\begin{aligned} \exists p\colon r'_p = m_p = 1 &\Leftrightarrow (n - m - w)_j = 1 \\ &\Leftrightarrow w_j = 1 \\ &\Leftrightarrow (w - m)_j = 0 \\ &\Leftrightarrow (w - m)_j \neq m_j \\ &\Leftrightarrow \nexists p\colon r_p = m_p = 1. \end{aligned}$$

For the last bit of part (ii) of Claim 2, we observe that the hypothesis for $n, m$ imply that we cannot have both $w < m$ and $n - w < m$ hold simultaneously. Thus, if $w < m$, we may still apply the first of the string of implications (87) above. Since $w_j = 0$ (else $w_j = 1$ and $w$ odd implies $w \geq m$), we have $r'_j = 0$, so both sides of (77) are odd. Similarly, if $n - w < m$, then $w_j = 1$ (else $w_j = 0$ and $w$ odd implies $n - m - w > 0$), then the string of implications (87) implies $r_j = 0$, so both sides of (77) are even. This completes the proof of the Proposition.

**Proof of Proposition (7.16)**

Write

$$|\psi_G\rangle = \frac{1}{\sqrt{2^n}} \sum_I c_I \, |I\rangle \tag{87}$$

$$|\psi_{D_1 G}\rangle = \frac{1}{\sqrt{2^{n-1}}} \sum_K d_k \, |K\rangle \tag{88}$$

$$|\psi_{S_1 G}\rangle = \frac{1}{\sqrt{2^{n-1}}} \sum_K s_k \, |K\rangle \tag{89}$$

so that we have

$$c_I = \prod_{e\in E}(-1)^{\wedge_e I} \tag{90}$$

$$d_K = \prod_{e:\,1\notin e}(-1)^{\wedge_e K} \tag{91}$$

$$s_K = \prod_{e\in E}(-1)^{\wedge_e K} \tag{92}$$

where we abuse notation to write $\wedge_e K$ to denote $\prod_{\ell\in e\setminus\{1\}} k_e$ for an $(n-1)$-bit string $K = k_2k_3\ldots k_n$. We can factor $c_I$ as

$$c_I = \prod_{e\in E}(-1)^{\wedge_e I} = \prod_{e:\,1\in e}(-1)^{\wedge_e I}\prod_{e:\,1\notin e}(-1)^{\wedge_e I} \tag{93}$$

so that we have

$$c_{0K} = \prod_{e:\,1\notin e}(-1)^{\wedge_e 0K} = d_K \tag{94}$$

$$c_{1K} = \pm\prod_{e:\,1\in e}(-1)^{\wedge_e 1K} = \pm s_K \tag{95}$$

where the sign in the last equation is $-1$ if $\{1\}$ is a hyperedge in $G$ and $+1$ otherwise. Thus we have

$$(\mathrm{tr}_1\,\rho_G)_{K,K'} = \frac{1}{2^n}\left(c_{0K}c_{0K'} + c_{1K}c_{1K'}\right) \tag{96}$$

$$= \frac{1}{2^n}\left(d_K d_{K'} + s_K s_{K'}\right) \tag{97}$$

$$= \frac{1}{2}\left((\rho_{D_1G})_{K,K'} + (\rho_{S_1G})_{K,K'}\right) \tag{98}$$

as desired.

**Proof of Proposition (7.17)**

Theorem 2.6 of [11] says there is a $2\times 2$ unitary matrix $\begin{bmatrix} a & b \\ c & d \end{bmatrix}$ such that

$$|\psi_H\rangle = a\,|\psi_{H'}\rangle + b\,|\psi_{K'}\rangle \tag{99}$$

$$|\psi_K\rangle = c\,|\psi_{H'}\rangle + d\,|\psi_{K'}\rangle\,. \tag{100}$$

It is elementary to check that the constraint that all coefficients in a hypergraph state vector are $\pm\frac{1}{\sqrt{2^n}}$ leads to just two possibilities $a = d = 1, b = c = 0$ and $a = d = 0, b = c = 1$.

**Proof of Proposition (7.18)**

By (7.17) and (7.16) we have that either $H = D_1G, K = S_1G$ or $H = S_1G, K = D_1G$. Applying (94) and (95) in the proof of (7.16) above, we have the desired result.